\documentclass[reprint,aps,prb,twocolumn,superscriptaddress,longbibliography]{revtex4-2}

\usepackage{amsmath,amssymb,amsfonts,bm,mathtools}
\usepackage{physics}
\usepackage{siunitx}
\usepackage{braket}
\usepackage{graphicx}
\usepackage{graphics,epsfig}
\usepackage{float}

\usepackage{xcolor}

\usepackage[hidelinks]{hyperref}

\begin{document}
	
	\title{Acoustic-phonon-driven spin–lattice relaxation of the hBN boron vacancy in the sub-THz regime}
	
	\author{Priyo Adhikary} 
    \email{ adhika38@purdue.edu}
	\affiliation{Elmore Family School of Electrical and Computer Engineering, Purdue University, West Lafayette, IN, USA.}
	\author{Pramey Upadhyaya}
		
	\affiliation{Elmore Family School of Electrical and Computer Engineering, Purdue University, West Lafayette, IN, USA.}
	\affiliation{ Purdue Quantum Science and Engineering Institute (PQSEI), Purdue University, West Lafayette, IN, USA.}

	\date{\today}
	
	\begin{abstract}
The negatively charged boron vacancy center in hexagonal boron nitride   is a premier candidate for quantum sensing, yet its performance is critically limited by longitudinal spin-lattice relaxation time ($T_1$). A microscopic understanding of spin relaxation in the high magnetic field regime remains elusive, as the relevant Zeeman transitions lie far below the optical phonon energies typically invoked to describe the relaxation process. Here, we apply an \textit{ab initio} acoustic mode spin-phonon relaxation theory to this problem and quantitatively reproduce the experimental magnetic field and temperature dependence of $T_1$ without empirical fitting parameters. We demonstrate that the relaxation dynamics are driven by a direct one-phonon emission and absorption process resonant with the Zeeman splitting. Furthermore, we identify the out-of-plane flexural phonon branch which is unique to two-dimensional hosts, as the primary source of decoherence, creating a distinct low-energy spectral function that facilitates spin relaxation. Our results provide a microscopic interpretation of the experimentally observed non-monotonic field and temperature dependence in two-dimensional quantum defect centers.
	\end{abstract}
	
	\maketitle
	
	\section{Introduction}
	The discovery of optically addressable quantum defect centers, most notably the nitrogen-vacancy (${\rm NV^{-}}$) center in diamond \cite{Schirhagl2014, DOHERTY20131}, has enabled a wide range of quantum technologies, particularly quantum sensing \cite{Degen17,Atature2018,Wolfowicz2021}, with applications including nanoscale magnetometry of condensed matter platforms \cite{Casola2018, Rovny2024}, thermometry \cite{Fujiwara_2021}, pressure sensing \cite{Ho2026}, and strain sensing \cite{Scholten21}. More recently, the negatively charged boron vacancy (${\rm VB^{-}}$) center in hexagonal boron nitride (hBN) has emerged as an attractive defect platform complementary to ${\rm NV^{-}}$ spin sensors, owing to its two-dimensional (2D) nature, which enables sensing in close proximity to target systems  \cite{Vaidya31122023,solanki2025,Durand23,Zhou24,Vlassiouk26, Huang2022, Gottscholl2021, Lyu2022}. Moreover, the van der Waals nature of hBN enables ${\rm VB^{-}}$ centers to be incorporated into ultrathin flakes that can be readily integrated with target platforms through well established van der Waals heterostructuring techniques, thereby forming near-surface spin defect layers. Furthermore, the ${\rm VB^{-}}$ center in hBN hosts an $S=1$ defect spin state, with spin density concentrated near the boron vacancy. Its spin quantization axis is naturally oriented perpendicular to the 2D plane, making it particularly well suited for extending quantum spin sensing into high-field, sub-THz environments \cite{solanki2025}.

    A central constraint on the performance of defect-based quantum sensors is the spin–lattice relaxation time ($T_1$). It sets the intrinsic limit on the spin polarization lifetime in relaxometry-based sensing schemes, and also bounds the achievable spin coherence time ($T_2$), thereby constraining overall sensing performance  \cite{Mondal2023}. Lattice vibrations of the host material play a central role in governing spin relaxation, making a first-principles predictive understanding of phonon induced spin dynamics in optically addressable spin defects critically important. Significant progress has been achieved toward this goal for both the ${\rm NV^{-}}$ and ${\rm VB^{-}}$ centers. However, to the best of our knowledge, existing first-principles treatments have largely focused on the near-room-temperature and low-field regime (typically $B \sim 1$ mT) \cite{Mondal2023}. In this limit, spin relaxation is typically dominated by two-phonon Raman processes mediated by relatively high-energy quasi-local optical phonon modes.

At high magnetic fields, however, the relevant spin transition energy increases with field, allowing a direct one-phonon process to become resonant with acoustic phonons. Although this mechanism is often neglected at low fields, it becomes essential once the spin splitting reaches the sub-THz regime \cite{solanki2025}. Capturing this physics microscopically requires Brillouin zone resolved spin–phonon coupling calculations that explicitly include finite-$\mathbf{q}$ acoustic phonons. Such \textit{ab initio} capabilities have only recently been demonstrated for bulk defect qubits, including the ${\rm NV^{-}}$ center \cite{gugler2018}, and for molecular qubits \cite{Lunghi19}. However, extending this formalism to truly 2D defect hosts remains comparatively underdeveloped, even though the flexural (ZA) phonon branch can qualitatively reshape the low-energy phonon bath and the resulting spin relaxation dynamics.

\begin{figure*}[htbp]
 	\centering
 	\includegraphics[width=\textwidth]{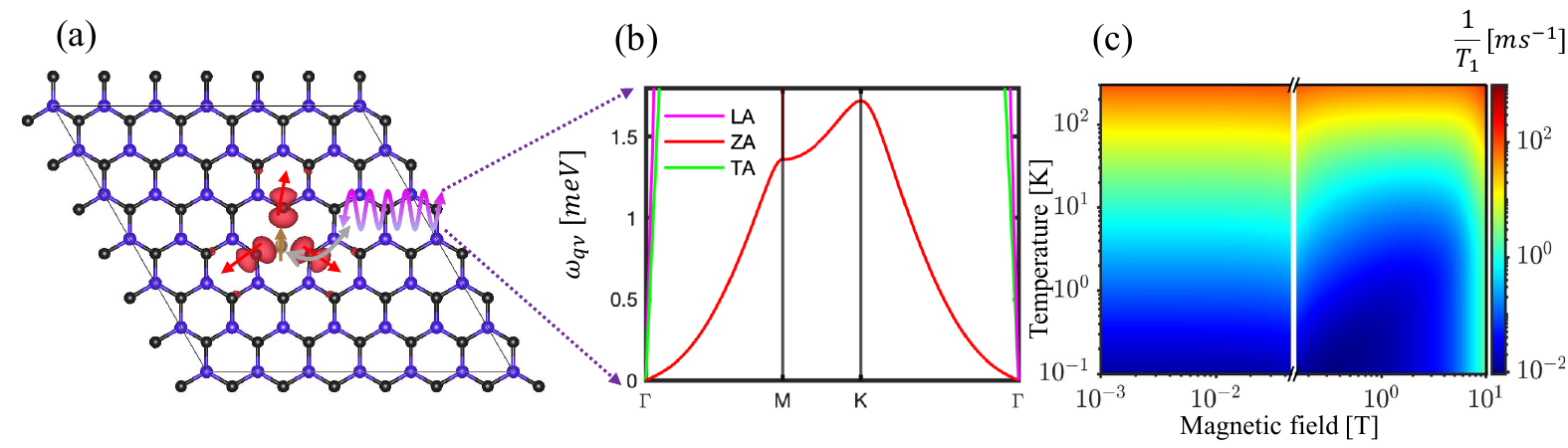}
 	\caption{Structural, vibrational, and spin relaxation overview of the ${\rm VB^{-}}$ center  in monolayer hBN.
 		(a)	Relaxed atomic structure of the  ${\rm VB^{-}}$ center  in a $6\times6\times1$ hBN supercell. Nitrogen atoms are shown in black and boron atoms in blue. The brown sphere with an arrow represents the vacancy site. The wavy line schematically depicts phonon coupled with the defect spin.  The red isosurface shows the magnetization density is concentrated on the three nearest-neighbor Nitrogen atoms of the defect spin $(S=1)$ as visualized using VESTA \cite{Momma2008}. (b) Low-energy acoustic phonon dispersion of the defect supercell along the high-symmetry path $\Gamma-\text{M}-\text{K}-\Gamma$, with the three acoustic branches color coded, longitudinal (LA, magenta), transverse (TA, green) and out-of-plane flexural (ZA, red). (c) Calculated spin lattice relaxation rate, $1/T_1$ of the ${\rm VB^{-}}$ center as a function of magnetic field  and temperature. The excluded region near the ground state level anticrossing field (${  B_{\text{cross}}} \approx 0.1$ T) indicates the unresolved region where spin transition energy falls below the  DFT phonon resolution limit.  
 		}
 	\label{fig:structure}
 \end{figure*}

Here, we fill this gap by extending finite-$\mathbf{q}$ spin–phonon calculations to the ${\rm VB^{-}}$ center in hBN within a fully parameter-free first-principles framework. We identify how low-energy acoustic phonon branches govern $T_1$ across experimentally relevant temperatures and magnetic fields, as summarized in Fig.~\ref{fig:structure}. In particular, we find that the direct one-phonon process gives rise to a pronounced magnetic field dependence at Tesla scale magnetic fields in qualitative agreement with recent experiments \cite{solanki2025}. Furthermore, we show that spin relaxation in the experimentally relevant field and temperature regime is dominated by the ZA  phonon branch. By providing a microscopic understanding of phonon mediated spin relaxation in 2D defect qubits across a broad range of magnetic fields and temperatures, our work offers key insights relevant to the development of high-field quantum sensing platforms based on layered materials.

 		\section{System Hamiltonian}

 		   \begin{figure*}[htbp]
 			\centering
 			\includegraphics[width=\textwidth]{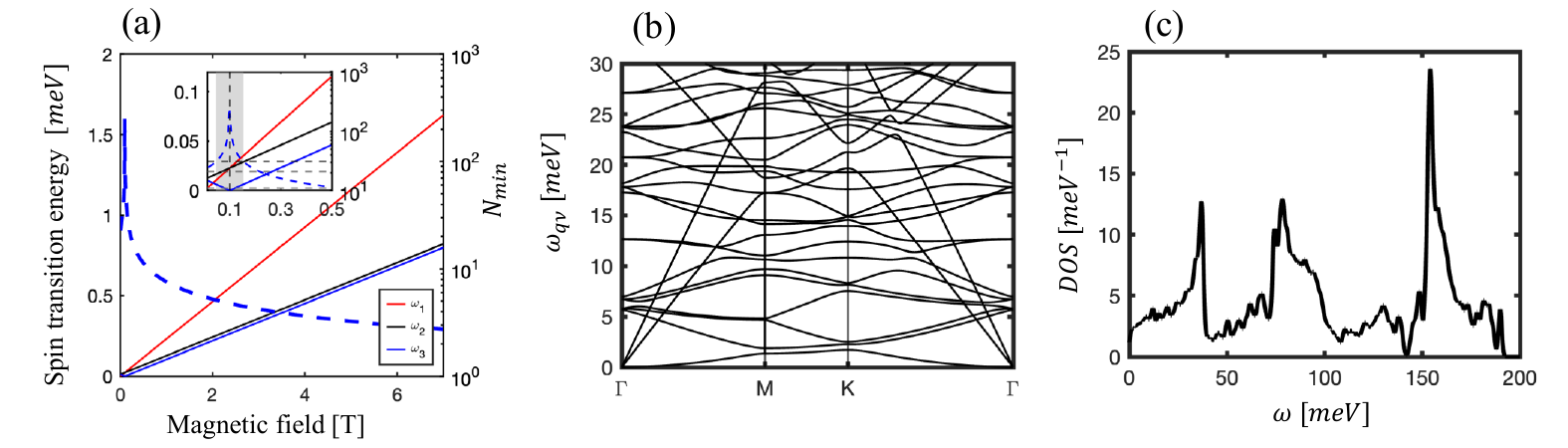 }
 			\caption{(a) Magnetic field dependence of the spin transition energies $\omega_1$, $\omega_2$, and $\omega_3$ (solid lines, left axis)  up to $B=7$ T. The blue dashed curve (right axis) shows the estimated minimum in-plane Brillouin zone mesh size ($N_{\rm min}$) as the transition energy  approaches zero.  Horizontal gray lines indicate reference mesh sizes of 11$\times$11, 21$\times$21 and 31$\times$31 used in the calculations of $T_1$.  The inset shows low-field region (  $B \le 0.5$ T), highlighting the ground state level anti-crossing at  ${  B_{\text{cross}}} \approx 0.1$ T (vertical dashed line), indicating that direct one-phonon relaxation near ${B_{\text{cross}}} $ requires very dense phonon sampling. The shaded region indicates the interval where the spin transition energy drops below the minimum phonon energy allowed for the DFT calculations. (b) Calculated phonon dispersion of the ${\rm VB^{-}}$ defect supercell along the high-symmetry path $\Gamma-\text{M}-\text{K}-\Gamma$, obtained within the harmonic approximation. (c)  Total phonon density of states at very low frequencies arises from the quadratic dispersion of the flexural ZA branch and provides the low-energy phonon bath relevant for the spin-phonon relaxation process.}
 			\label{fig:transition}
 		\end{figure*}

The dynamics of the ${\rm VB^{-}}$ center are governed by interactions between the defect spin and lattice vibrations. The total Hamiltonian is given by the sum of the static spin Hamiltonian, the phonon bath Hamiltonian, and the spin-phonon interaction Hamiltonian \cite{Mondal2023,Cambria23,gugler2018,Norambuena18,Lunghi19,Liu25,Estaji25}. The static spin Hamiltonian, including the Zeeman interaction induced by an external magnetic field $\mathbf{B}$, is
\begin{equation}
	\label{eq:hamiltonian_S}
	H_{\rm spin}
	=
	\sum_{ij} S_i D_{ij} S_j
	+
	\gamma_s \mathbf{B}\cdot\mathbf{S},
\end{equation}
where $i$ and $j$ denote the Cartesian components ($x,y,z$) of the defect spin operator $\mathbf{S}$, and $D_{ij}$ is the zero-field-splitting (ZFS) tensor. For the ${\rm VB^{-}}$ center, ZFS tensor is traceless, and the scalar parameter ($D$) entering the  transition frequencies ($\omega_2$, $\omega_3$) follows the standard convention ($D=\frac{3}{2} D_{zz}$ )\cite{Norambuena18,Mondal2023}.  

The lattice dynamics are treated within the harmonic approximation and described by phonon modes with frequencies $\omega_{\mathbf{q}\nu}$ and creation (annihilation) operators $b^{\dagger}_{\mathbf{q}\nu}$ ($b_{\mathbf{q}\nu}$),
\begin{equation}
	H_{\rm p}
	=
	\sum_{\mathbf{q}\nu}
	\hbar\omega_{\mathbf{q}\nu}
	\left(
	b^{\dagger}_{\mathbf{q}\nu} b_{\mathbf{q}\nu}
	+
	\frac{1}{2}
	\right),
\end{equation}
where $\mathbf{q}$ and $\nu$ label the phonon wavevector and branch, respectively.
 
Spin–phonon coupling arises from the modulation of the ZFS tensor by lattice vibrations. Phonon induced atomic displacements modify the defect’s local crystal field environment and thereby changes the ZFS tensor. We treat these distortions as a weak perturbation and expand it in a Taylor series in the phonon normal mode coordinates ($\mathbf{R}$),
  \begin{align}
	D_{ij}(\{ \bf  R\}) &\approx  D^{(0)}_{ij}
	+ \sum_{\mathbf{q}\nu} \left( \frac{\partial D_{ij}}{\partial R_{\mathbf{q}\nu}} \right)_0 R_{\mathbf{q}\nu}\nonumber \\ &
	+ \frac{1}{2} \sum_{\mathbf{q}\nu} \left( \frac{\partial^2 D_{ij}}{\partial R_{\mathbf{q}\nu}^2 } \right)_0 R_{\mathbf{q}\nu}^2  +O(R^3).
\end{align}
 The subscript $0$ indicates that the derivatives are evaluated at the relaxed (equilibrium) structure and $D^{(0)}$ is the corresponding  ZFS tensor.
 Truncating the series up to second order yields the effective spin-phonon interaction Hamiltonian,

\begin{align}
	V^{(1)} &= \sum_{\mathbf{q}\nu}\sum_{ij} \left( \frac{\partial D_{ij}}{\partial R_{\mathbf{q}\nu}} \right)_0 R_{\mathbf{q}\nu} \, S_i S_j 	\label{eq:hamiltonian_r}\\[1em]
	V^{(2)} &= \frac{1}{2} \sum_{\mathbf{q}\nu} \sum_{ij} \left( \frac{\partial^2 D_{ij}}{\partial R_{\mathbf{q}\nu}^2} \right)_0 R_{\mathbf{q}\nu}^2   \, S_i S_j , \label{eq:hamiltonian_r_2}
\end{align}
 where  $V^{(1)} $ and $		V^{(2)} $ are the first and second order spin–phonon coupling terms respectively. 


\section{Computational methods}
	
	The static spin levels and phonon spectrum of the ${\rm VB^{-}}$ center are determined from the ZFS tensor and phonon eigenmodes, respectively, obtained within spin-polarized density functional theory (DFT) \cite{Barth_1972, Hobbs2000, Steiner2016}. The phonon-induced spin relaxation rate is then obtained by applying Fermi's golden rule to the spin-phonon coupling term evaluated via DFT \cite{VanVleck1940, Lunghi2023, Cambria23}. To this end, we performed spin-polarized DFT calculations on a $6\times6\times1$ monolayer hBN supercell containing a single boron vacancy~\cite{Estaji25,Viktor2026}. We adopted the $6\times6\times1$ supercell on the basis of earlier convergence studies, which compared this cell size with a $12\times12\times1$ supercell for the ${\rm VB^{-}}$ center and found negligible finite-size effects on $T_1$ \cite{Estaji25}. Fig.~\ref{fig:structure}(a) shows the relaxed structure of the ${\rm VB^{-}}$ center.  Calculations were carried out using the Vienna \textit{Ab initio} Simulation Package (VASP) \cite{vasp1} with the projector augmented wave (PAW) method, and exchange correlation was treated within the generalized gradient approximation (GGA) using the Perdew–Burke–Ernzerhof (PBE) functional \cite{vasp2}. For Brillouin zone sampling we use a $\Gamma$-centered $4\times4\times1$ $k$ mesh with a plane wave kinetic energy cutoff of 520\,eV.  We use  20\,\AA{} vacuum spacing  along the $c$-axis to eliminate spurious interlayer interactions. Electronic self-consistency was converged to $10^{-7}$ eV, and structural relaxations were continued until the residual forces were below  $10^{-4}$\,eV/\AA.


	Phonon properties were computed within the harmonic approximation using the finite displacement approach as implemented in the \textsc{Phonopy} package \cite{TOGO20151, phonopy-phono3py-JPCM, phonopy-phono3py-JPSJ}. We generated symmetry inequivalent displaced supercells, evaluated the Hellmann–Feynman forces using the same DFT settings, and constructed the second order force constant (Hessian) matrix.  Diagonalization of the dynamical matrix yields phonon eigenfrequencies (\(\omega_{\mathbf{q}\nu}\)) and   normalized eigenvectors  (\(\mathbf{e}_{\mathbf{q}\nu}\)). These mode resolved quantities are then used to build finite-$\mathbf{q}$  spin–phonon couplings. To this end, we evaluate the gradient terms in  Eq.~(\ref{eq:hamiltonian_r}), (\ref{eq:hamiltonian_r_2}) via a finite difference scheme.
For the second-order contributions, we use a diagonal approximation, neglecting the computationally expensive off-diagonal mode-mixing terms \cite{Cambria23,Estaji25}. Specifically, we neglect mixed derivatives of the form $(\partial^2 D_{ij}/\partial R_{q\nu}\partial R_{q'\nu'} $ with $q\nu \neq q'\nu')$. No analogous approximation is required for the first-order coupling, which is linear in a single phonon coordinate.
	The coupling coefficients are extracted by distorting the atomic positions along each phonon polarization direction by a magnitude of $\Delta Q = 0.1 \, \text{\AA} \sqrt{{\rm amu}} $ \cite{Cambria23}.  For a detailed derivation of the finite difference scheme, see Appendix~\ref{app:finite_diff}. 
	 
	Previous studies have established that accurate determination of the ZFS tensor requires hybrid functionals \cite{Heyd2003} combined with spin-contamination corrections \cite{ Liu25,Ivady2020}.  However, this approach is computationally expensive for calculating ZFS on a dense phonon mesh. Alternative methods using cluster embedding have been proposed in which a finite region (7.5 {~\AA} radius) is extracted from the supercell, hydrogen passivated, and treated using a quantum chemistry method \cite{ORCA}, while phonons are computed for the periodic system \cite{Mondal2023, Liu25}. While this approximation is justified by the localized nature of spin-phonon coupling, it relies on different structural models for the electronic and vibrational calculations.  We employ a unified single supercell framework to compute both the ZFS tensor and phonon properties.
We computed the  ZFS  of the ${\rm VB^{-}}$ center within VASP using the native spin–spin dipolar formalism \cite{vaspzfs}, obtaining $D= $2.76 GHz. While the absolute ZFS may differ from experiment, spin–phonon relaxation rates are governed primarily by the changes in the ZFS tensor under atomic displacements. Following \cite{Estaji25}, we assume that any systematic offset in the absolute $D$  value is similar for the equilibrium and displaced geometries and therefore largely cancels in the ZFS derivatives that enter the spin–phonon coupling.

\section{Results}


 \subsection{Spin transitions and \textit{ab initio}  phonon spectrum}
The ground state of the ${\rm VB^{-}}$ center, as described by the static spin Hamiltonian in Eq.~\eqref{eq:hamiltonian_S}, consists of an $S=1$ triplet manifold. The corresponding energy-level separations are characterized by three transition frequencies. The frequency $\omega_1=\gamma_s B$ corresponds to the Zeeman splitting between the $m_s=-1$ and $m_s=+1$ states. The remaining transition frequencies are $\omega_2=D+\gamma_s B$ and $\omega_3=\lvert D-\gamma_s B\rvert$, corresponding to the transitions between $m_s=0$ and $m_s=+1$, and between $m_s=0$ and $m_s=-1$, respectively. Fig.~\ref{fig:transition}(a) shows the evolution of these transition energies for magnetic fields up to $B=7~\mathrm{T}$. A ground-state level anticrossing occurs at $B_{\rm cross}=D/\gamma_s$, where $\omega_3$ vanishes.

Fig.~\ref{fig:transition}(b) shows the phonon dispersion of the ${\rm VB^{-}}$ center along the high-symmetry path $\Gamma$--M--K--$\Gamma$. The large supercell leads to band folding and a corresponding increase in the number of phonon branches relative to pristine monolayer hBN. Nevertheless, the gapped optical branches and the gapless acoustic character of the low-energy modes remain clearly identifiable \cite{hbnphonon}. In particular, two acoustic branches (LA and TA) exhibit linear dispersion, while the flexural ZA branch displays the characteristic quadratic dispersion $\omega \sim \sigma_{\rm ZA}q^2$ near the $\Gamma$ point. Here, $\sigma_{\rm ZA}$ is the flexural coefficient. The corresponding phonon density of states (DOS), shown in Fig.~\ref{fig:transition}(c), remains finite at low frequencies owing to the quadratic dispersion of the ZA mode. The physical nature of these modes is illustrated in Fig.~\ref{fig:phononvect}, where the phonon polarization vectors \cite{Roy2024} reveal the in-plane and out-of-plane atomic displacements that couple to the defect spin.

\begin{figure}[htbp!]
	\centering
	\includegraphics[width=\columnwidth]{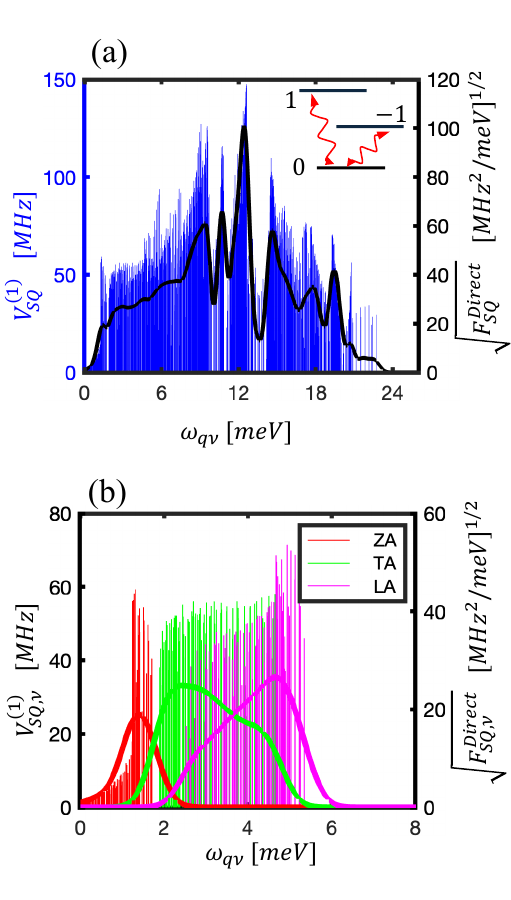}
	\caption{First order spin-phonon coupling of the ${\rm VB^{-}}$ center  for single quantum transitions (SQ).  (a)   Discrete vertical lines denote spin-phonon coupling for discrete  phonon frequencies, and the solid curve shows the corresponding spectral function. Inset: schematic of the direct one-phonon process driving the SQ transition, with the wavy line denoting the emitted/absorbed phonon resonant with the spin transition energies. (b) $V^{(1)}_{SQ,\nu}$ represents the first order mode-resolved spin-phonon coupling for the three low-energy acoustic branches: ZA (red), TA (green), and LA (magenta). Vertical lines denote the discrete coupling magnitude for the branch selected at each phonon momenta $\mathbf{q}$, obtained from the first-principles calculations. Solid curves show the corresponding branch resolved spectral functions.  The ZA branch dominates the low-energy (\(\hbar\omega_0\lesssim 0.8~\mathrm{meV}\)) spin relaxation.}
	\label{fig:firstorder}
\end{figure}

We note that there is a large mismatch between the optical phonon energies and the spin-transition energies. Optical modes therefore do not contribute to resonant one-phonon spin relaxation. In contrast, acoustic phonons can induce resonant one-phonon spin transitions over the experimentally relevant field range, $0 \le B \le 7~\mathrm{T}$ \cite{solanki2025}, where the ground-state transition frequencies span $\sim 0$--$200~\mathrm{GHz}$ ($\hbar\omega_3 \lesssim 0.8~\mathrm{meV}$). In particular, the flexural ZA branch, which exhibits a finite low-frequency phonon DOS, can contribute significantly to spin relaxation. Accurately capturing these resonant one-phonon processes therefore requires resolving the full phonon dispersion of the defect supercell, including acoustic branches extending to very low energies.

To quantify the Brillouin zone sampling required to resolve the resonant ZA phonons, we estimate the minimum mesh size, $N_{\min}$, by matching the spin-transition energy to the ZA dispersion. Using the field dependence of the transition frequency $\omega_3$, the resonant wave vector is given by $q(B)=\sqrt{\omega_3(B)/\sigma_{\rm ZA}}$. The corresponding minimum mesh size is then estimated as $N_{\min}(B)\approx b_{\rm mag}/q(B)$, where $b_{\rm mag}$ denotes the magnitude of the in-plane reciprocal lattice vector. For a fixed Brillouin zone mesh of size $N_q$, the smallest nonzero wave vector is approximately $q_{\min}\approx b_{\rm mag}/N_q$, corresponding to a minimum resolvable phonon energy $\omega_{\rm ZA}(q_{\min})$. Consequently, the resonance condition cannot be fully resolved within the field interval $|B-B_{\rm cross}| < \omega_{\rm ZA}(q_{\min})/\gamma_s$ around the level anticrossing. For $N_q=31$, this unresolved window spans approximately $0.045$--$0.153~\mathrm{T}$. The corresponding field range is indicated by the shaded region in the inset of Fig.~\ref{fig:transition}(a) and in subsequent figures.

\begin{figure}[htbp]
	\centering
	  \includegraphics[width=\columnwidth]{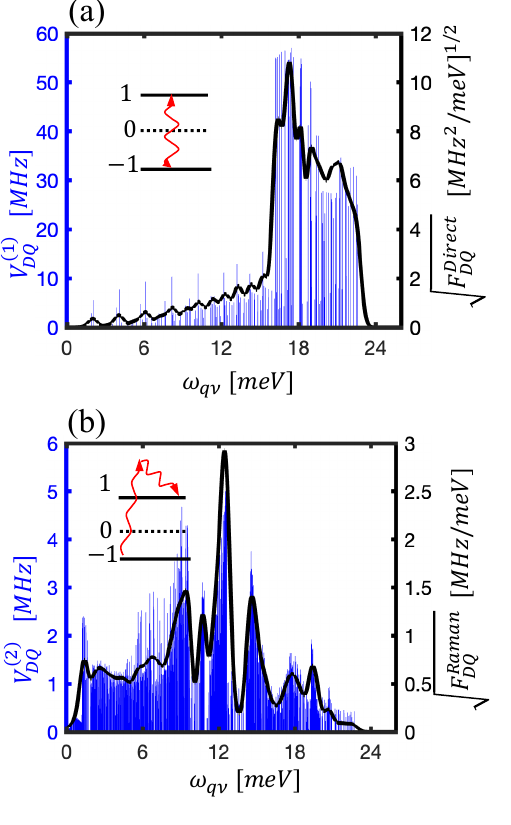}
	\caption{ Spin-phonon coupling  for double quantum transition (DQ). (a) First order spin–phonon coupling, shown as discrete vertical lines, while solid curve represents the corresponding spectral function. Inset: schematic of the direct one-phonon process driving the DQ transition, with the wavy line denoting the emitted/absorbed phonon resonant with the spin transition energy. (b) Second order spin–phonon coupling, shown as discrete vertical lines and the solid curve representing the spectral function of the ${\rm VB^{-}}$ center. Inset: schematic of the two-phonon Raman process for the DQ transition with the wavy line denoting the emitted  and absorbed phonon between two spin states.  The SQ transition is forbidden at the second order because of the $D_{3h}$ selection rule of the ${\rm VB^{-}}$ center \cite{Mondal2023,Estaji25}.  }
	\label{fig:secondorder}
\end{figure}
 
\subsection{Mode resolved spin-phonon couplings and spectral functions}

We calculate the spin-phonon coupling for each phonon mode and wave vector from the change in the ZFS tensor of the ${\rm VB^{-}}$ center induced by the corresponding phonon normal-mode displacement. The first order spin-phonon interaction, $V^{\rm (1)}(\mathbf{q}\nu)$, is defined by the product of the spin bilinear operator and the gradient of the ZFS tensor, as given in Eq.~\eqref{eq:hamiltonian_r}. We project this interaction onto distinct magnetic transition sectors, namely the single-quantum (SQ) transitions ($m_s=0 \rightarrow m_s=\pm1$) and the double-quantum (DQ) transition ($m_s=-1 \rightarrow m_s=1$).

Figs.~\ref{fig:firstorder}(a) and \ref{fig:secondorder}(a) show the resulting spin-phonon couplings for the SQ and DQ transitions, respectively. Comparison of the two channels reveals that the SQ coupling is stronger than the DQ coupling. Furthermore, the mode-resolved spin-phonon couplings associated with the ZA, LA, and TA branches are overlaid as discrete vertical lines in Fig.~\ref{fig:firstorder}(b). The results demonstrate that, within the experimentally relevant energy range ($\lesssim 1~\mathrm{meV}$), the spin-phonon coupling is dominated by the flexural ZA phonons, whereas the contributions from the LA and TA branches become appreciable only at higher energies. 

To characterize the frequency dependence of the spin-phonon interaction induced relaxation, we calculate the spectral function. For each magnetic transition channel (SQ, DQ), the spectral function for the direct process is defined as \cite{Norambuena18,Lunghi2023},
	\begin{equation}
	\label{eq:Fdef}
	F^{Direct}(\omega) = \frac{1}{N_q} \sum_{\mathbf{q}\nu} \left| V^{(1)}(\mathbf{q}\nu) \right|^{2} \delta(\omega - \omega_{\mathbf{q}\nu}),
\end{equation}
where $N_q$ is the size of the Brillouin zone. 
We map the discrete spin-phonon couplings obtained from DFT calculations to a continuous spectral function using a Gaussian broadening function ($\rho_{\sigma}$). See Appendix~\ref{app:optimization} for a detailed discussion on the optimal broadening parameter ($\sigma$). The spectral function incorporates the effects of both the phonon DOS and the strength of the spin-phonon coupling. We show the spectral functions for the first order process for  single  and double quantum transitions as solid black lines in Fig.~\ref{fig:firstorder}(a) and Fig.~\ref{fig:secondorder}(a) respectively. The spectral function for single quantum transitions is significantly larger than that of the double quantum transitions, particularly in the low-energy regime. The mode-resolved spectral function for the single-quantum transition, shown in Fig.~\ref{fig:firstorder}(b), reveals that the low-energy contribution arises predominantly from the ZA modes. 
We applied a frequency cutoff of 26 meV to ensure the inclusion of all three acoustic (LA, TA and ZA) dispersions. This truncation is justified because the direct relaxation process involves resonance with the spin splitting, which operates at an energy scale $\le 1$ meV. Consequently, higher-energy optical phonons are energetically decoupled from the spin transitions and do not contribute to the direct relaxation rates.

  The quadratic term in Eq.~\eqref{eq:hamiltonian_r_2} defines the second order spin-phonon coupling ($V^{(2)}(\mathbf{q}\nu)$). We project this interaction onto the same magnetic transition sectors as in the first order analysis. The spectral function for the second order process is defined similarly to that in Ref. \cite{Cambria23, Estaji25} as,
        \begin{align}
  	\label{eq:Fdef}
  	F^{Raman}(\omega) &= \frac{1}{N_q} \int d\omega' \sum_{\mathbf{q}\nu } \left| V^{(2)}(\mathbf{q}\nu) \, \delta(\omega' - \omega_{\mathbf{q}\nu})\right|^{2} \nonumber \\
  	&\quad \times \frac{1}{\sigma\sqrt{2\pi}} \exp\left(-\frac{(\omega'-\omega)^2}{2\sigma^2}\right)
  \end{align}
    We include only the second-order coupling in the Raman channel because the  alternative two-phonon process mediated by two successive first-order 
  interactions is  suppressed by the large virtual-state energy denominator \cite{Cambria23}. 
  In Fig.~\ref{fig:secondorder}(b) we show the second order couplings and the corresponding spectral function. The results imply that the double quantum channel dominates over the single quantum channel at second order, consistent with earlier $ \Gamma$-point optical phonon calculations on ${\rm VB^{-}}$   center \cite{Estaji25}.  We include finite momentum phonons in the calculation and find that this trend persists, rather than being restricted to $ \Gamma$-point optical phonons.

	\subsection{Spin–lattice relaxation rate }
	
	The total longitudinal spin relaxation rate is written as a sum of first and second order spin–phonon contributions,
	\begin{equation}
		  \frac{1}{T_1}= 	\Gamma_{\mathrm{Direct}}(B,T) + \Gamma_{\text{Raman}}(T)
	\end{equation}
 The direct term describes single-phonon processes resonant with the Zeeman tuned spin splittings, while the Raman term captures two-phonon emission and absorption processes \cite{Mondal2023,Estaji25}. This decomposition is consistent with experimental phenomenology, wherein the high-field regime exhibits a characteristic magnetic field dependence indicative of a resonant one-phonon channel, whereas the high temperature behavior is well described by a weakly field-dependent two-phonon mechanism \cite{solanki2025}. 
 
 We compute $ \Gamma_{\mathrm{Direct}} $  by summing the contributions of the three allowed transitions between the triplet sublevels, using $F^{Direct}(\omega)$ discussed in the previous section. The rate is given by \cite{Norambuena18},
 		\begin{equation}
				\label{eq:firstorder1}
		\Gamma_{\mathrm{Direct}}(B,T)
		=\sum_{j=1}^{3}F^{Direct}(\omega_j)  	\coth\left( \frac{\omega_j(B)}{2k_B T} \right) ,
	\end{equation}   
with the thermal factor equivalently written as  $2n_B(\omega_j,T) + 1$, in terms of the Bose occupation,
	\begin{equation}
		n_B(E,T)=\frac{1}{\exp\left(E/k_BT\right)-1}.
		\label{eq:Bose}
	\end{equation}
	The three transition frequencies ($\omega_j$) depend on the ZFS  and the Zeeman energy. This leads to a sharp peak in the thermal occupation factor near  $B_{\text{cross}}$. For $\omega << k_B T$, the thermal factor can be approximated as, 
	$\coth\left( \frac{\omega(B)}{2k_B T} \right) \approx \frac{k_B T}{\gamma_{s}\abs{B-{ B_{\text{cross}}}}}$.  In Fig. \ref{fig:spinphncoup2}(b) we  plot the magnetic field dependence of the Bose occupation factor, which shows the sharp enhancement as $B$  approaches  $B_{\text{cross}}$. 
	We can understand this behavior qualitatively by noting that as $B$  approaches  $B_{\text{cross}}$, the spin transition energy tends to zero. In this limit the spectral function is approximately constant. Therefore,  the field dependence of $\Gamma_{\mathrm{Direct}}(B,T)$ near  $B_{\text{cross}}$ is governed primarily by the Bose occupation factor. Furthermore, this also implies that a peak in $1/T_1$ at $B_{\text{cross}}$ persists at all temperatures. As we move away from $B_{\text{cross}}$ at higher fields, $\Gamma_{\mathrm{Direct}}(B,T)$ increases with $B$. This is because the resonant frequencies shift into a regime where $F^{Direct}(\omega)$ grows rapidly with $\omega$ (see Fig.~\ref{fig:firstorder}), thereby strengthening the one-phonon contribution. 

 Using the spectral function defined in Eq.~\eqref{eq:Fdef}, the relaxation rate for the two-phonon (Raman) absorption–emission process is given by~\cite{Cambria23},
 
   	\begin{equation}
		 \Gamma_{\text{Raman}} = \frac{2\pi}{\hbar} \int  d\omega  F^{Raman}(\omega)  n_B(\omega, T) \left[ n_B(\omega, T) + 1 \right] .
		 \end{equation}      
This is a non-resonant process that activates only when the energy difference between two-phonons matches the spin splitting. As shown in earlier works \cite{Norambuena18, solanki2025}, this contribution is weakly dependent on the magnetic field and depends strongly on temperature. 
	 	 \begin{figure*}[htbp]
		\centering
		\includegraphics[width=\textwidth]{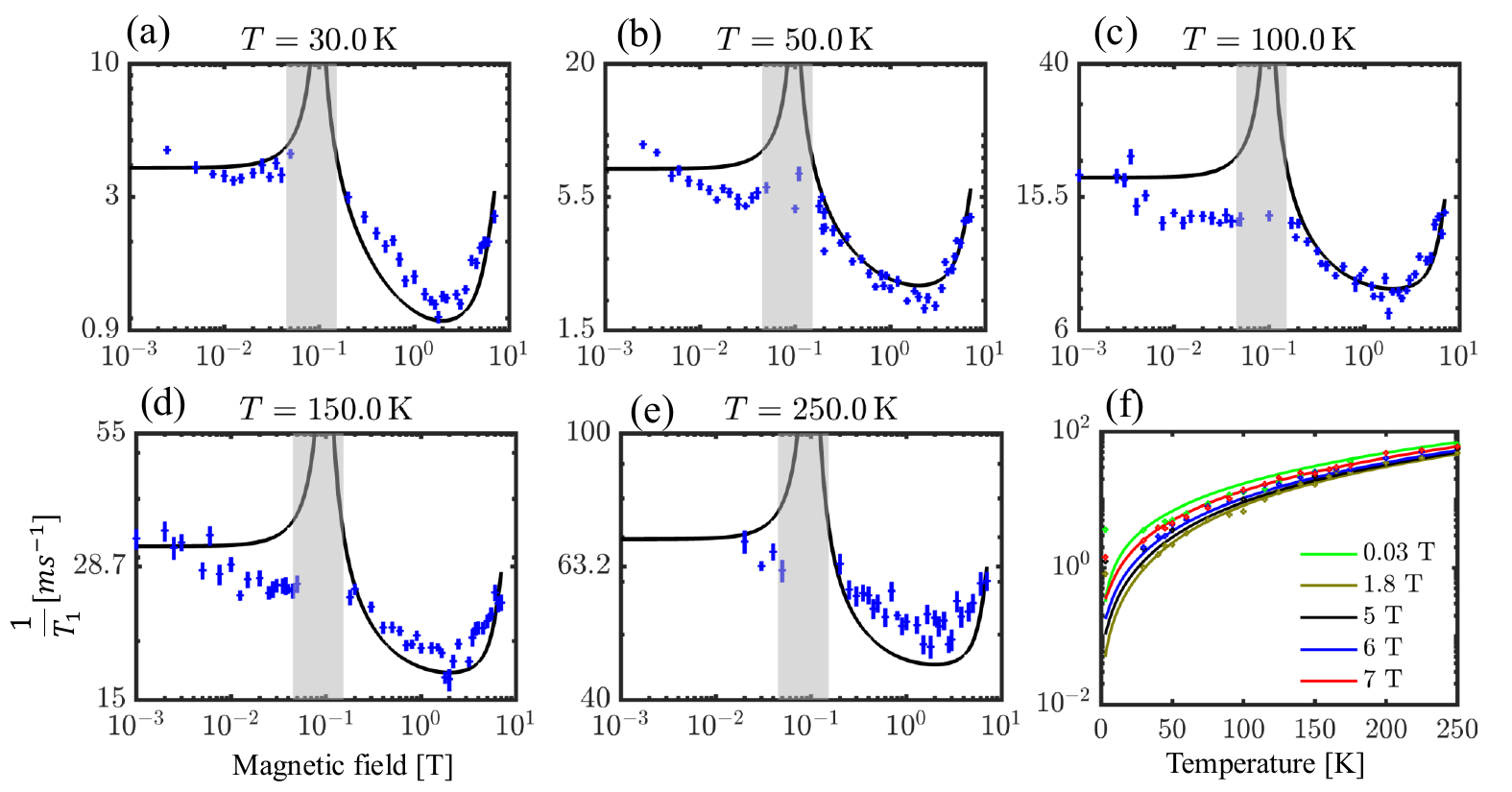}
		\caption{ Magnetic field and temperature dependence of the spin-lattice relaxation rate $\frac{1}{T_1}$ of  the ${\rm VB^{-}}$ center. (a)-(e) Field dependence of $\frac{1}{T_1}$  at $T= 30, 50, 100, 150, 250 $ K. Experimental data are shown as blue points \cite{solanki2025} and solid lines denote the calculated total rate.  (f) Temperature dependence of $\frac{1}{T_1}$  at fixed magnetic fields $B=0.03, 1.8, 5, 6, 7$ T.   The shaded region highlights the field interval in which the spin transition energy falls below the DFT resolution limit. }
		\label{fig:ratefiled}
	\end{figure*}

    In Fig.~\ref{fig:structure}(c) we summarize the main result of this work: ${T_1}$ as a function of magnetic field and temperature using the DFT-derived spin-phonon coupling. The overall non-monotonic field dependence and the temperature scaling reflect the interplay between the resonant direct one-phonon process and the two-phonon process. 
	Furthermore, in Figs.~\ref{fig:ratefiled}(a)–(e), we plot the calculated  $1/T_1$ as a function of magnetic field for temperatures ranging from  30 to 250 K, with the experimental data overlaid from Ref. \cite{solanki2025}.  We use  \textit{ab initio} results to show that the inclusion of the direct one-phonon channel is required to reproduce the high-field relaxation. As the Zeeman tuned splitting approaches $\sim 200$ GHz, low-energy spin relaxation proceeds via resonant one-phonon absorption and emission. 
	
	 In earlier phenomenological analyses of the experimental data, this regime was identified through a temperature-weighted power-law dependence on the magnetic field. By explicitly treating the acoustic phonons within a first-principles framework, we provide a microscopic interpretation of this field dependence and directly relate it to the frequency dependence of the one-phonon spectral function evaluated at the spin-transition energies.

We also note deviations between the experimental measurements and the calculated $T_1$ values, particularly in the lower-field regime. These differences may originate from two main sources. First, the phonon spectral function in experiment may differ from that of the pristine monolayer considered here, since the measurements are performed on a thicker hBN flake supported on a substrate. Second, spin relaxation may involve additional channels beyond the purely phonon-mediated mechanism captured in the present calculations. For example, cross relaxation arising from dipolar interactions with nearby paramagnetic defects may provide an additional contribution to the relaxation rate, analogous to the behavior commonly observed in ${\rm NV^{-}}$ ensembles in nanodiamonds \cite{Guillebon20}. Quantifying such effects requires going beyond the present \textit{ab initio} framework and is therefore left for future work.

	 Next, we examine the temperature dependence of   $1/T_1$  at fixed magnetic fields ranging from 0.03 to 7 T, as illustrated in Fig.~\ref{fig:ratefiled}(f) alongside the experimental data.  The relaxation rate is not monotonic in magnetic field. The rate at 0.03 T exceeds that at the intermediate field of 1.8 T, despite the systematic increase at high fields from 1.8 to 7 T. Within the \textit{ab initio} framework, this behavior follows because the direct one-phonon rate is governed by two competing factors evaluated at the spin transition frequencies, $\coth\left( \frac{\omega_j(B)}{2k_B T}\right)$ and $F^{Direct}(\omega)$.  As the magnetic field increases, the transition frequency moves out of the thermally enhanced low-energy window, reducing the thermal factor, but the spectral function simultaneously increases sharply at higher frequencies, producing the net increase in the relaxation rate at high-fields.  At high temperatures, the two-phonon Raman process provides the dominant contribution to the relaxation rate. In this regime, the calculated $1/T_1$  approaches $T^2$-like dependence, consistent with earlier \textit{ab initio} work based on $\Gamma$-point optical phonon analysis \cite{Mondal2023,Estaji25}. Upon lowering temperature, $1/T_1$  at different magnetic fields separate progressively because the direct one-phonon process becomes more important than the weakly field-dependent two-phonon process.

	\section{Summary}
	
	For the bulk ${\rm NV^{-}}$ center, the spin–phonon coupling has  been introduced through a phenomenological ansatz, which was subsequently invoked to rationalize the well-known  $T^{5}$ temperature scaling of the spin-lattice relaxation rate \cite{Norambuena18}. In contrast, for the ${\rm VB^{-}}$ center, \textit{ab initio} calculations~\cite{Mondal2023,Estaji25} and previous numerical fits~\cite{solanki2025} point to a different low temperature behavior, with $1/T_1 \propto T^2$. This distinct temperature scaling originates from phonon modes intrinsic to the 2D nature of the hBN lattice.
	
	We present a first-principles theory for 2D quantum defects that unifies the magnetic field and temperature dependence of the spin relaxation rate.  In particular, accounting for finite momentum acoustic phonons provides the relevant low-energy (\(\hbar\omega_0\lesssim 0.8~\mathrm{meV}\)) bath required to describe the experimentally accessed splittings across $0 \leq B \leq 7$~T \cite{solanki2025}. More broadly, the ZA-mediated low-energy phonons constitute a qualitatively different relaxation landscape from bulk defects, and it provides a route to interpreting and engineering decoherence in 2D platforms.
	
	In conclusion, we develop an acoustic phonon spin-relaxation framework for the ${\rm VB^{-}}$ center in monolayer hBN by computing finite momentum, mode resolved spin-phonon couplings from first-principles and constructing the corresponding first and second order spectral functions. This approach yields a parameter-free description of $1/T_1$ across temperature and magnetic field as the combined contribution of direct one-phonon and Raman-like  processes. By moving beyond a $\Gamma$-point optical phonon approximation, the theory identifies acoustic modes as the dominant decoherence channel at high-fields and provides a predictive basis for tailoring relaxation via phonon engineering in 2D quantum defects.

	\textit{Acknowledgements:}
    The authors acknowledge Prof. Vladimir M. Shalaev, Prof. Benjamin Lawrie, Abhishek Bharatbhai Solanki and Aravindh Shankar for helpful discussions.  The authors also thank Dr.  Abhishek Bharatbhai Solanki  for sharing the experimental data. The theoretical analysis was supported by National Science Foundation under Award No. 1944635. The authors also acknowledge support from the W. M. Keck Foundation grant “Uncovering Elusive Kitaev Topological States with Metasurfaces”. 
    This work used computational resources provided by the Rosen Center for Advanced Computing at Purdue University, including the Gautschi community cluster.
    
	\makeatletter
	\nocite{apsrev4-2Control}
	\bibliography{references}
	
	\appendix  
		  \begin{figure}[htbp]
		\centering
		\includegraphics[width=\columnwidth]{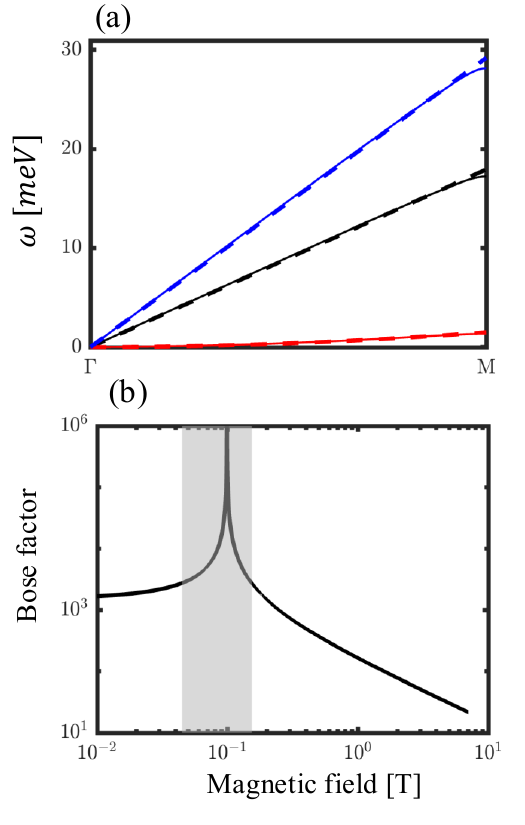}
		\caption{(a) low-energy acoustic phonon dispersions near $\Gamma$ fitted to $\omega_{\rm LA}=v_{\rm LA} q$ (blue) , $\omega_{\rm TA}=v_{\rm TA} q$ (black) , $\omega_{\rm ZA}=\sigma_{\rm ZA} q^2$ (red). Dashed lines show the fitted dispersions, while solid lines indicate the corresponding first-principles phonon bands.  (b)  Magnetic field dependence of the thermal Bose occupation factor and spin-phonon spectral function for the  ${\rm VB^{-}}$ center.  }
		\label{fig:spinphncoup2}
	\end{figure}
	\section{Finite difference derivatives along  phonon modes}
	
	\label{app:finite_diff}
	To evaluate the dependence of ZFS on lattice vibrations, we obtain the gradients of the ZFS tensor $\mathbf{D}$ with respect to the phonon normal modes using a finite difference approach.  We use  \textsc{Phonopy} package~\cite{TOGO20151} to obtain phonon eigenvectors and generate distorted structures by displacing atoms along selected normal modes.
	For a given mode $\nu$ at wavevector $\mathbf{q}$, the cartesian displacement $\Delta{\bf R}_{i,\nu}$ of atom $i$ was derived from the mass weighted normalised eigenvector $\mathbf{e}_{i, \nu}$ using,
	\begin{equation}
		\Delta \mathbf{R}_{i, \nu} = \pm \Delta Q \frac{\mathbf{e}_{i, \nu}}{\sqrt{M_i}},
	\end{equation}
	where $M_i$ is the atomic mass and $\Delta Q$ is a fixed numerical displacement amplitude set to $0.1 \, \text{\AA}\sqrt{\text{amu}}$ \cite{Cambria23,Estaji25}. This uniform scaling factor was chosen to ensure the structural distortions remained within the harmonic regime. For a given phonon mode,  we displaced the nuclei by $\pm\Delta Q$ along the normal coordinate $R_{\mathbf{q\nu}}$ and evaluated $\mathbf{D}^{+}\equiv \mathbf{D}( {+}\Delta Q)$, $\mathbf{D}^{-}\equiv \mathbf{D}({-}\Delta Q)$.  
	\begin{figure*}[htbp]
		\centering
		\includegraphics[width=\textwidth]{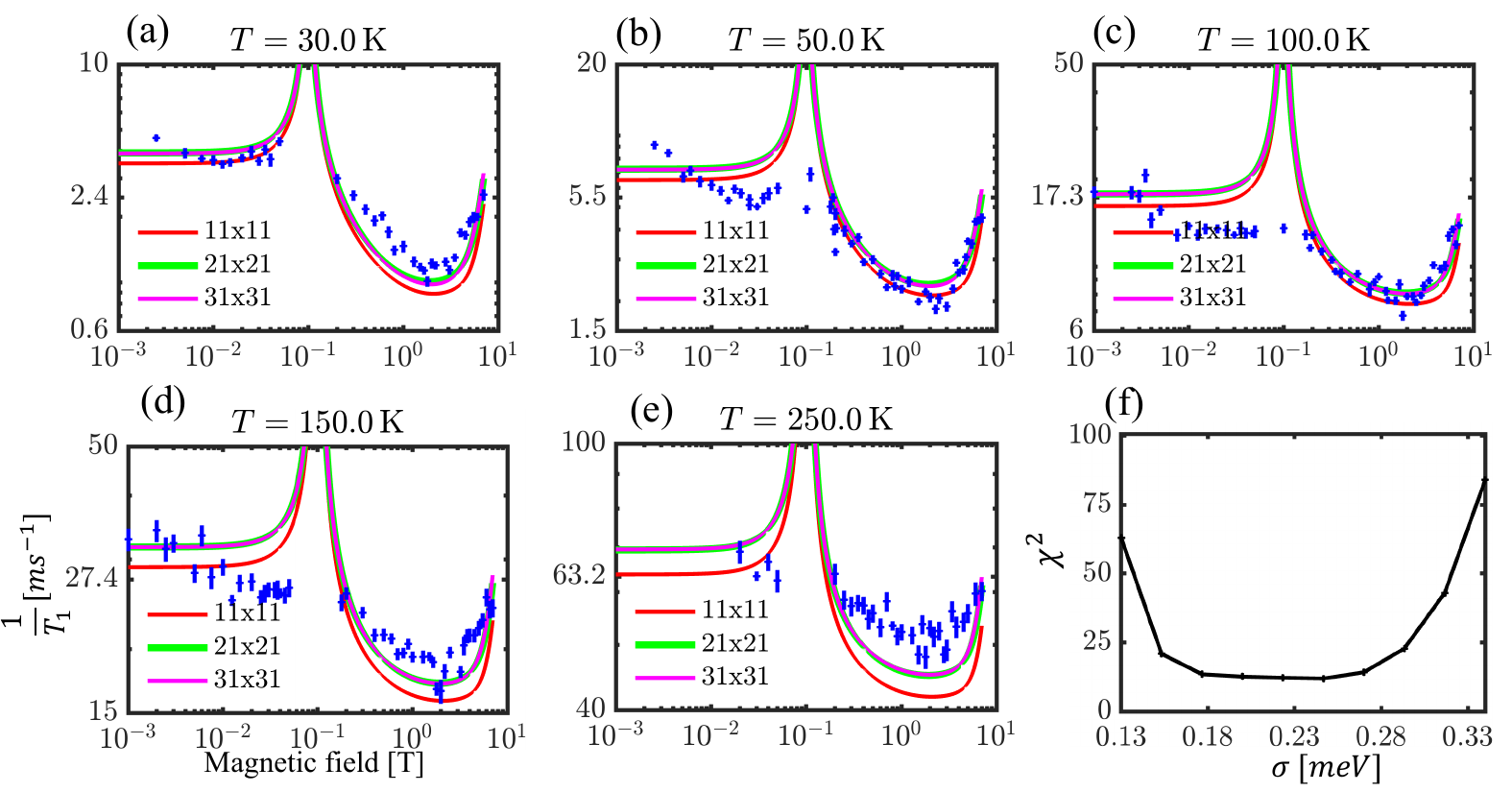}
		\caption{ (a)-(e) Comparison of the experimental and calculated spin-lattice relaxation rate $\frac{1}{T_1}$  as a function of magnetic field at temperatures of 30, 50, 100, 150, and 250 K respectively. Experimental data are shown as blue points. Solid lines represent first-principles calculations performed on phonon q-point meshes of $11\times11$, $21\times21$, and $31\times31$.  The results from the $21\times21$ and $31\times31$ meshes are well-converged at low temperatures.  (f) Optimization of the Gaussian broadening parameter $\sigma$. The $\chi^2$ deviation between the experimental data and the theoretical rates (calculated for the $31\times31$ {\bf q}-mesh) is plotted against $\sigma$.}
		\label{fig:error}
		\end{figure*}

	We obtain, 
	\begin{align}
		\label{eq:FD1}
		\nabla\mathbf{D}_{\mathbf{r}} 
		&= \frac{\mathbf{D}^{+} - \mathbf{D}^{-}}{2\,\Delta Q},\\[4pt]
		\label{eq:FD2}
		\nabla^{2}\mathbf{D}_{\mathbf{r}} 
		&= \frac{\mathbf{D}^{+} + \mathbf{D}^{-} - 2\mathbf{D}^0}{(\Delta Q)^2}.
	\end{align}
	Equations \eqref{eq:FD1} and \eqref{eq:FD2} approximate the linear and quadratic response of $\mathbf{D}$ to $Q_{\mathbf{q\nu}}$, respectively. $\mathbf{D}^0$ is the ZFS for relaxed atomic geometry.

	 \begin{figure*}[htbp]
	\centering
	\includegraphics[width=\textwidth]{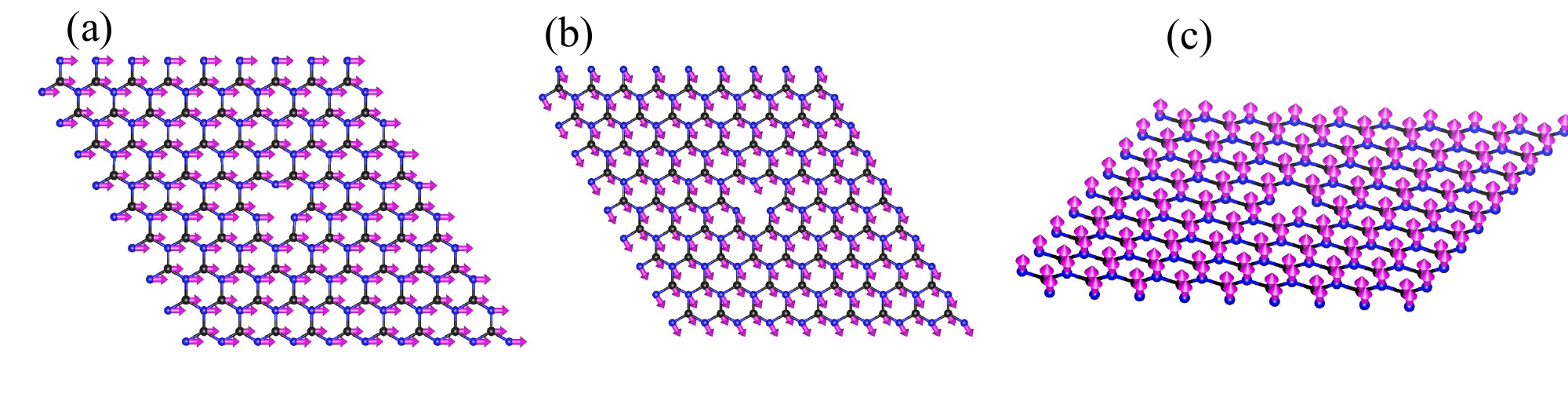}
	\caption{ Visualization of lowest-energy acoustic phonon modes for ${\rm VB^{-}}$ center. The magenta arrows represent phonon eigenvectors, indicating the direction of the atomic displacements. (a) and(b) are in-plane acoustic modes and (c) depicts out-of-plane acoustic mode.}
	\label{fig:phononvect}
\end{figure*}

	\section{Phonon energies and Gaussian kernel}

    \subsection{Debye model parameterization of the low-energy acoustic phonons}
	\label{app:debye}
	To obtain an analytic description of the \textit{ab initio} spectral function,  we use the Debye model, parameterized from the DFT acoustic dispersions for the ${\rm VB^{-}}$ center. This yields an approximate  phenomenological description of the spin–phonon coupling in the sub-THz regime.  The fitted dispersion in the vicinity of the $\Gamma$ point is shown in Fig.~\ref{fig:spinphncoup2}(a) along with the DFT results.  
	We assumed linear dispersion, $\omega_{\mathrm{LA/TA}}(q) \approx v_{\mathrm{LA/TA}}\,q$, for the in-plane modes and quadratic dispersion, $\omega_{\mathrm{ZA}}(q) \approx \sigma_{\mathrm{ZA}} q^{2}$, for the flexural branch. This procedure gives in-plane sound velocities of $v_{\mathrm{LA}}=18.46$~km/s and $v_{\mathrm{TA}}=11.33$~km/s, along with a flexural coefficient of $\sigma_{\mathrm{ZA}}=3.95\times10^{-7}$~m$^2$/s. These fitted parameters define an effective phonon DOS and enable a clear separation of low-energy acoustic contributions from the optical modes in the subsequent analytic reconstruction of the spin–phonon coupling. 
    
	\subsection{Optimization of the Gaussian broadening parameter}
	\label{app:optimization}
Energy conservation for the direct spin transition implies the presence of a Dirac delta function in the spin–phonon spectral function.  
However, first-principles DFT calculations yield a discrete phonon spectrum  $\{\omega_{q\nu}\}$ because the Brillouin zone integral is evaluated on a finite ${\bf q}$ mesh of size $N_q$. The presence of the Dirac delta function produces a numerically ill-conditioned Brillouin zone sum.  To bridge the disparity between the discrete ${\bf q}$-point sampling and the continuum limit, we replace the singular Dirac delta with a   normalized Gaussian kernel, 
\begin{equation}
	\label{eq:gauss}
	\delta(\Delta) \equiv \rho_{\sigma}(\Delta) = \frac{1}{\sigma\sqrt{2\pi}} \exp\left(-\frac{\Delta^2}{2\sigma^2}\right)
\end{equation}
where $\sigma$ is the broadening width. Accordingly, we approximate the delta function by a regular function that enables integration of the spectral function over the Brillouin zone.
This regularization also implies that convergence must be achieved using both (1) the  broadening width and (2) Brillouin zone sampling size. 
We define reduced $\chi^2$as,  
\begin{equation}
	\chi^2(\sigma)=\frac{1}{N-1}\sum_{k=1}^{N}
	\left[
	\frac{T_{1,\rm Th}^{-1}(B_k,T_k;\sigma)-T_{1,\rm exp}^{-1}(B_k,T_k)}
	{\delta\!\left(T_{1,\rm exp}^{-1}\right)_k}
	\right]^2,
	\label{eq:chi2_rate}
\end{equation}
To determine the optimal $\sigma$, we performed a reduced $\chi^2$ minimization analysis comparing the theoretical spin-lattice relaxation rates to experimental data \cite{solanki2025} across  available magnetic field and temperature range.  We use Gaussian widths in the range $\sigma \in [0.10, 0.30]$ meV, for the optimization procedure. The resulting $\chi^2$ is shown in Fig.~\ref{fig:error}(f).  We choose  $\sigma =0.25$ meV which provides a faithful numerical representation of the Dirac delta constraint while maintaining stable convergence across the different ${\bf q}$ mesh sampling procedures used in our rate calculations.\\

\subsection{Mesh size dependence of  $T_1$ }
	\label{app:mesh}
In Fig.~\ref{fig:error}(a)-(e) we illustrate the \textit{ab initio} spin-phonon relaxation rates  as a function of the  magnetic field for  temperature range of $30-250$ K. The comparison of phonon ${\bf q}$  meshes, $11\times11$, $21\times21$, and $31\times31$, reveals that $T_1$ is dependent on the Brillouin zone sampling size. $T_1$ calculated on the $11\times11$ mesh,  deviates from the results obtained with denser meshes. The rate calculated on $21\times21$ and $31\times31$ mesh shows minimal variation, demonstrating convergence has been achieved.  We adopt the $31\times31$  mesh for all subsequent calculations of spin-phonon couplings and relaxation rates of the ${\rm VB^{-}}$  center.

\end{document}